\documentclass[aps,prd,twocolumn,groupedaddress,showpacs]{revtex4}
\usepackage{graphicx}
\usepackage{amsmath}
\begin{document}

\title{Strongly Coupled Chameleons and the Neutronic Quantum Bouncer}

\author{Philippe Brax}
\affiliation{Institut de Physique Th\'eorique, CEA, IPhT, CNRS, URA 2306,
  F-91191Gif/Yvette Cedex, France}
\email{philippe.brax@cea.fr}

\author{Guillaume Pignol}
\affiliation{LPSC, Universit\'e Joseph Fourier, CNRS/IN2P3, INPG, Grenoble, France}
\email{guillaume.pignol@lpsc.in2p3.fr}

\date{\today}

\begin{abstract}
We consider the potential detection of chameleons using bouncing ultracold neutrons. We show that the presence of a chameleon field
 over a planar plate would alter the energy levels of ultra cold neutrons in the terrestrial gravitational field.
When chameleons are strongly coupled to nuclear matter, $\beta\gtrsim 10^8$, we find that the shift in energy levels would be detectable with the forthcoming GRANIT experiment, where
a sensitivity of  order one percent of a peV is expected. We also find that an extremely large coupling $\beta\gtrsim 10^{11}$ would lead to new bound states at a distance of order 2 microns, which is already ruled out by previous Grenoble experiments. The resulting  bound, $\beta\lesssim 10^{11}$, is already three orders of magnitude better than the upper bound, $\beta\lesssim 10^{14}$, from precision tests of atomic spectra.
\end{abstract}

\maketitle

The accelerated expansion of the Universe is  one of the most puzzling discoveries of observational cosmology  so far.
Amongst all  the various attempts to explain this  phenomenon, such as the existence of a pure cosmological constant justified by anthropic considerations, a possible modification of gravity on large scales or the fact that the Copernican principle could be violated, the existence of a cosmological scalar field of the quintessence type \cite{Copeland:2006wr,caldwell,trodden,Brax:2009ae} is a strong contender just  as Cold Dark Matter is  a good candidate to explain the rotation curves of galaxies and the large scale structures of the Universe.
Recently, models in which scalar fields couple to matter \cite{KhouryWeltman} have been developed and led to the motivated possibility of testing them in laboratory experiments.
Indeed, a  scalar field coupled to ordinary matter might  mediate a new long range force and would show up in  fifth force searches or equivalence principle tests.
In order to evade the resulting  constraints, screening mechanisms have been invoked whereby the field surrounding a compact body becomes trapped inside and does not lead to a large gradient outside  dense objects.
In this letter, we will focus on the chameleon mechanism where
a combination of the potential $V(\phi)$  of the scalar field $\phi$ and a coupling to matter leads to the existence of an effective potential for the scalar field quanta which depends on the local density $\rho$ of the environment
\begin{equation}
V_{\rm eff}(\phi)= V(\phi) +e^{\beta \phi/M_{\rm Pl}} \rho.
\label{veff}
\end{equation}

So far the best constraint on $\beta$ comes from atomic physics and reads $\beta \lesssim 10^{14}$\cite{clare}.
When the potential is of the runaway type with a vanishing minimum at infinity,
the effective potential has a density dependent minimum
 $\phi_{\rm min}$.
This is the vacuum of the theory in a given environment.
The  density-dependent minimum is such that  the mass
of the scalar field becomes also density dependent. We will focus  on inverse power law models defined by
\begin{equation}
V(\phi)= \Lambda^4+ \frac{\Lambda^{4+n}}{\phi^n}+ \cdots
\label{veff2}
\end{equation}
where we have neglected higher inverse powers of the chameleon field.
We will choose $\Lambda=2.4 \times 10^{-12} \, {\rm GeV}$ to lead to the acceleration of the universe on large scales.
The potential has a minimum located at
\begin{equation}
\phi_{\rm min}= \left( \frac{n M_{\rm Pl} \Lambda ^{4+n}}{\beta
\rho} \right)^{1/(n+1)}.
\end{equation}
The chameleon rest  mass at the minimum is
\begin{equation}
m_\phi^2\approx \beta \frac{\rho}{M_{\rm Pl}} \frac{n+1}{\phi_{\rm min}} .
\label{mass}
\end{equation}
Thus, when considering a macroscopic body of size $d$ as a source of the field $\phi$, nonlinear effects of the theory become significant when $m_\phi d\gtrsim 1$.
In this case, only a thin shell at the surface of the body contributes to the field.
The thickness of the shell (and the very presence of the thin shell when considering small bodies) depends on the strength $\beta$ of the chameleon coupling to matter.
Due to the thin shell effect, the chameleon could very well be strongly coupled and still evade laboratory limits with macroscopic bodies \cite{mota}.


When searching for a fifth force, one can study the interaction between two macroscopic bodies (torsion pendulum, Casimir force...) or one can use subatomic particles.
Macroscopic bodies have a better sensitivity for a macroscopic range of  the force, for example the Seattle Eotwash experiment is the best probe for millimeter range fifth forces \cite{EotWash}.
Experiments studying the Casimir effect are competitive probes of extra interactions in the micrometer range.
They could also test chameleon models despite the presence of the thin shell effect \cite{Casimir}.
Subatomic particles are probing shorter ranges. For example, neutron scattering data provide the best constraints at the nanometer scale \cite{Nesvizhevsky:2007by}.

In addition, there are experiments looking for the possible interaction of subatomic particles with a macroscopic body.
In this article we consider the experiments probing  ultracold neutrons bouncing above a mirror and analyse their sensitivity to the chameleon field.
We expect a measurable effect as the neutron will not display a thin-shell effect due to the large extension of its wave function in the terrestrial gravitational field above the mirror. The resulting shift in the neutron energy levels induced by the unscreened change in the potential energy of neutrons due to the chameleon could be detectable.
In the following we will calculate the chameleon field produced by the mirror used in the bouncing neutron experiments, and then investigate how the neutron could probe this chameleon field on top of the mirror.

We first consider the chameleon profile when a dense plate is embedded in a sparse environment.
We consider an infinitely thick plate $z\in ]-\infty,0]$ of density $\rho$ in contact with a  vacuum for $z\ge 0$ of density $\rho_\infty$.
The chameleon profile satisfies the Klein-Gordon equation
\begin{equation}
\frac{d^2\phi}{dz^2}= V'_{\rm eff}(\phi)
\end{equation}
both inside and outside the plate where $V_{\rm eff}$ is given by eq. (\ref{veff}) and (\ref{veff2}).
We assume that both deep inside and at infinity, the chameleon settles at an effective minimum satisfying
\begin{equation}
V'_{\rm eff}(\phi_{b,\infty})=0
\end{equation}
where $\phi_b$ and $\phi_\infty$ are respectively the bulk and infinity vacua.
In the following we shall assume that $\beta \phi / M_{\rm Pl} \ll 1$ implying that
\begin{equation}
\frac{\beta}{M_{\rm Pl}} \rho_{b,\infty}= -V'(\phi_{b,\infty}).
\end{equation}
Therefore we find that  in the bulk and outside the plate
\begin{eqnarray}
\frac{d^2\phi}{dz^2} &= V'(\phi)- V'(\phi_b) & \quad {\rm (bulk)} \\
\frac{d^2\phi}{dz^2} &= V'(\phi)- V'(\phi_\infty) & \quad {\rm (outside)}.
\end{eqnarray}
Integrating these equations determines  the  value of $\phi_0=\phi(0)$:
\begin{equation}
\phi_0= \frac{ V(\phi_b) -V(\phi_\infty) -\phi_b V'(\phi_b) + \phi_\infty V'(\phi_\infty)}{V'(\phi_\infty) -V'(\phi_b)}
\end{equation}
and therefore
\begin{equation}
\sqrt{2} z=\frac{\phi_\infty^{n/2+1}}{\Lambda^{2+n/2}}\int_{x_0}^x \frac{y^{n/2} dy}{ (1- y^n + n(y^{n+1}- y^n))^{1/2}}
\end{equation}
where $\phi>\phi_0$ when $z>0$
and
 $x= \phi/\phi_\infty$.
As long as $x\ll 1$ we can approximate the integrand by $x^{n/2}$ and therefore
\begin{equation}
z\approx \frac{\sqrt 2}{2+n} \frac{\phi_0^{n/2 +1}}{\Lambda^{2+ n/2}} \left( \left(\frac{\phi}{\phi_0} \right)^{n/2+1}-1 \right)
\end{equation}
Now the mass of the chameleon at the boundary is
\begin{equation}
m_\phi^2 \sim n(n+1) \frac{\Lambda^{4+n}}{\phi_0^{n+2}}
\end{equation}
and we find for the field profile
\begin{equation}
\label{generalprofile}
\phi= \phi_0 \left( 1+ \frac{(2+n)}{\sqrt {2 n(n+1)}} m_\phi z \right)^{2/(2+n)}.
\end{equation}
When the densities are $\rho_\infty\ll \rho$ we have $\phi_\infty \gg \phi_b$ and
\begin{equation}
\phi_0 \approx \frac{\phi_b}{n} \approx \frac{1}{n} \left( n \frac{\Lambda^{n+4}M_{\rm Pl}}{\beta \rho} \right)^{1/(n+1)}.
\end{equation}

The contribution of the chameleon to the interaction potential with matter particles of mass $m$ is $\beta m/M_{\rm Pl} \phi$ implying that the total potential
gets an extra attractive term in addition to the usual gravitational acceleration $g=9.8$~m/s$^2$:
\begin{equation}
\label{pot1}
\Phi(z) = m gz + \beta \frac{m}{M_{\rm Pl}} \phi(z).
\end{equation}

Let us now analyse the phenomenological consequences of the presence of the additional potential above the mirror.
We consider ultracold neutrons bouncing above the mirror, with a mirror density of $\rho = 2.6$~g/cm$^3$ ($10^{19}$~eV$^4$ in natural units) to be specific.
The vertical motion of the bouncing neutrons is described by the stationary Schr\"odinger equation for the wavefunction $\psi$:
\begin{equation}
\label{schrodinger}
-\frac{\hbar^2}{2 m} \frac{d^2}{dz^2} \psi + \Phi(z) \psi(z) = E \psi(z), \quad \psi(0) = 0,
\end{equation}
where $m$ is the mass of the neutron.
From now on we do not work in natural units anymore and reintroduce the $\hbar$ factors.
As a generic feature of a quantum particle confined in a well, the energy $E$ can only assume discrete values $(E_k)_{k=1,2 \cdots \infty}$.
In absence of the chameleon effect, the quantum bouncer problem (\ref{schrodinger}) can be solved exactly in terms of the Airy function ${\rm Ai}$ and its negative zeros ${\rm Ai}(-\epsilon_n) = 0$:
\begin{equation}
\label{solutionSchrodinger}
E_k = E_0 \, \epsilon_k \ ,  \ \psi_k(z) = c_k {\rm Ai}(z/z_0 - \epsilon_k) \ {\rm for} \ z \geq 0
\end{equation}
where $c_k$ is a normalization factor, $E_0 = mg z_0 = 0.6$~peV 
and $z_0$ is the (very large) spatial extension of the wavefunctions:
\begin{equation}
z_0 = \left( \frac{\hbar^2}{2 m^2 g} \right)^{1/3} = 5.87 \, \mu {\rm m}
\end{equation}
The unusual scale of the parameters of the quantum states is due to the weakness of the gravitational force, it makes this quantum system very sensitive to additional interactions.
Following the experimental discovery of the quantization of the energy levels of the bouncing neutrons at the ILL reactor in Grenoble \cite{nature},
an extensive programme has been started in order to probe this phenomenon with increased precision \cite{Sanuki,abele,Kreuz} (a recent review on the topic can be found in \cite{ReviewQuantumStates}).
In particular, the forthcoming GRANIT experiment being set up at the ILL \cite{Kreuz} will measure the energy levels with a precision better than $0.01$~peV.
Ultimately even better sensitivity could be reached, down to $10^{-7}$~peV \cite{abele2}.
We will now argue that the neutron bouncer is a competitive probe of the chameleon field in the case of strongly coupled chameleons,
since the additional term in (\ref{pot1}) leads to two potentially observable effects: the shrinking of the wavefunctions of the stationary states and the shifting of the energy levels.

\begin{figure}
\centering
\includegraphics[width=0.87\linewidth,angle=90]{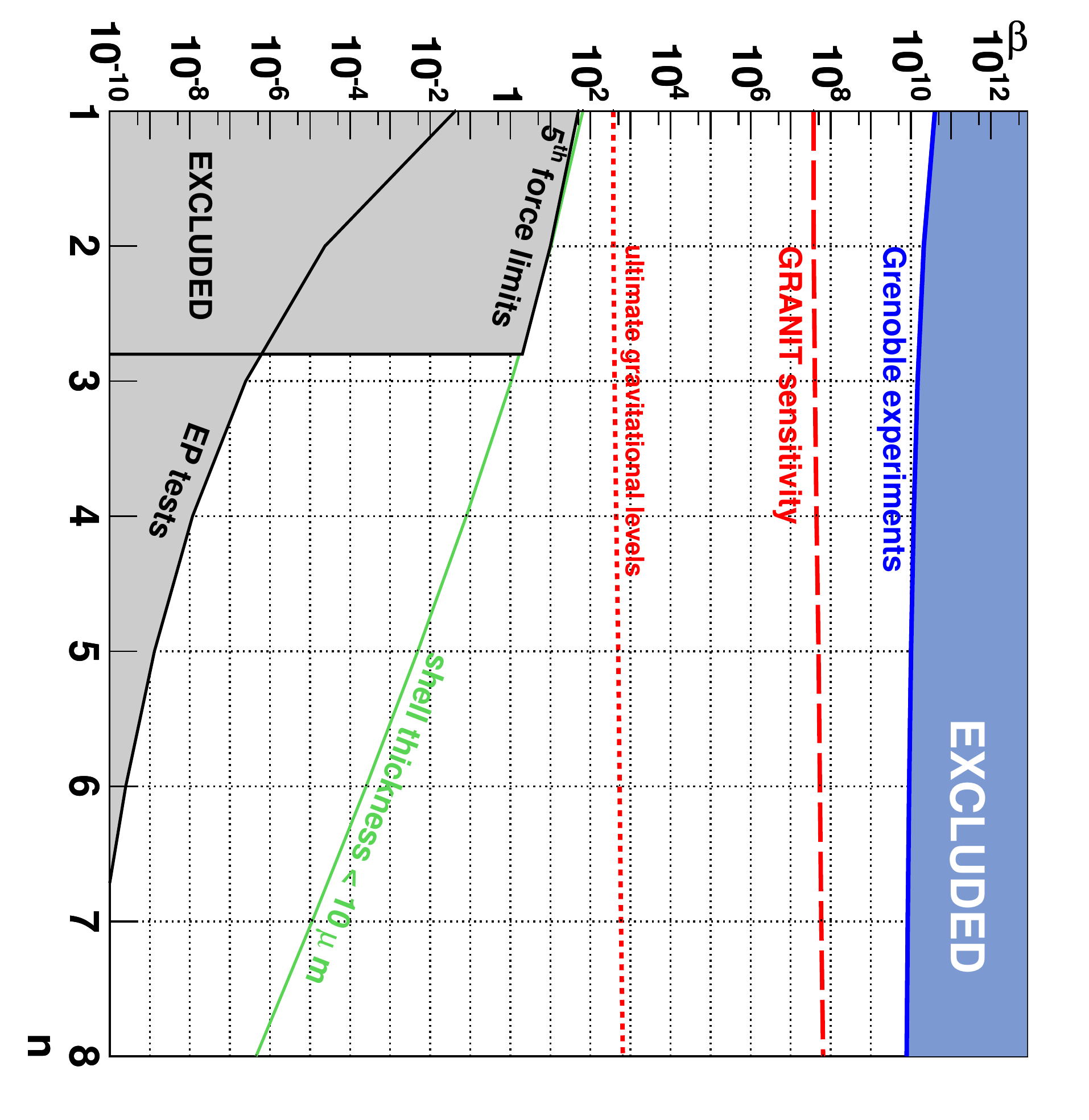}
\caption{The chameleon exclusion plot.
We find that above the  bottom green line the  chameleon field is independent of the coupling,
above the top blue line  chameleons produce a quantum state with a size of 2 microns and finally
above the red dashed line the chameleons shift the $3 \rightarrow 1$ resonance by more than $0.01$~peV. 
We have also drawn the ultimate sensitivity limit at the $10^{-7}$ peV level. 
}
\label{exclusion}
\end{figure}

In the case where the chameleon is strongly coupled $\beta \gg 1$, we find that $m_\phi z_0 \ll 1$ and the chameleon field (\ref{generalprofile}) seen by the ultracold neutrons a few micrometers above the mirror
simplifies:
\begin{equation}
\label{phiSimple}
\phi(z) = \Lambda \left( \frac{2+n}{\sqrt{2}} \Lambda z \right)^{2/(2+n)}.
\end{equation}
The line in the parameter space $(\beta, n)$ where $m_\phi z_0 = 1$ is shown as a green line in fig. \ref{exclusion}, justifying the simplification (\ref{phiSimple}).
This blue line corresponds to the situation where the mirror has a thin shell effect of thickness $z_0$.
Importantly, in this regime, the chameleon field $\phi(z)$ above the surface is independent of the coupling strength $\beta$.
This explains why experiments with macroscopic bodies have no net  gain in sensitivity for large couplings:
when $\beta$ increases, the thin shell at the surface of the test bodies shrinks and  there is no increase in the force.
The situation is drastically different for neutron gravitational quantum states.
Although the mirror has a thin shell the neutron has none,
the net potential seen by the neutron is linear in $\beta$
(it would be quadratic in $\beta$ in the absence of the thin shell effect for the mirror).
Also, it is important to notice that the chameleon field is independent of the density of the mirror.

To study quantitatively the phenomenology of the chameleon term, it is useful to write the  potential in the following form:
\begin{equation}
\Phi(z) = mgz + \beta V_n (z/\lambda)^{\alpha_n}
\end{equation}
with
\begin{eqnarray*}
V_n & = & \frac{m}{M_{\rm Pl}} \Lambda  \left( \frac{2+n}{\sqrt{2}} \right)^{2/(2+n)} \\
  & = & 0.9 \times 10^{-21} \ {\rm eV} \ \left( \frac{2+n}{\sqrt{2}} \right)^{2/(2+n)} \\
\end{eqnarray*}
where  $\lambda  =  1/\Lambda = 82 \ \mu {\rm m}$ and $\alpha_n   =  2/(2+n)$.
Besides the Planck mass, the chameleon field introduces a single new characteristic distance scale $\lambda$, which is remarkably close to the size of the bouncing neutron wavefunctions.

Now let us study two effects of the additional potential, i.e. a  modification of the wavefunction of the neutron  and a shift in the energy levels.
If strong enough, the attractive new term $\beta V_n (z/\lambda)^{\alpha_n}$ could by itself create an additional bound state very close to the surface
and would have shown up in early experiments performed at ILL Grenoble.
One can estimate the size of the first quantum state from the Heisenberg relation.
For a classical bouncing motion of turning point $Z$, the momentum oscillates with maximal value
\begin{equation}
p_{\rm max} = \sqrt{2m \beta V_n} (Z/\lambda)^{\alpha_n/2}.
\end{equation}
Imposing now the Heisenberg relation  $p_{\rm max} Z = \hbar$ gives the characteristic height of the ground state level:
\begin{equation}
Z = \left( \frac{\hbar^2 \lambda^{\alpha_n}}{2m \beta V_n} \right)^{1/(\alpha_n+2)}.
\end{equation}
In order not to conflict with the Grenoble experiments, the height $Z$ must be larger than about $Z_{\rm lim} = 2 \, \mu$m.
This kind of approach has already been used to set constraints on short range attractive Yukawa forces \cite{Nes}.
In our case we find the limit:
\begin{equation}
\beta < \frac{\hbar^2}{2m} \frac{\lambda^{\alpha_n}}{V_n Z_{\rm lim}^{\alpha_n+2}}
\end{equation}
which is plotted in Fig. \ref{exclusion}.

\begin{table}
\begin{center}
\begin{tabular}{l|llllllll}
n 		& 1     & 2     & 3     & 4     & 5     & 6     & 7     & 8    \\
$\alpha_n$	& $2/3$ & $1/2$ & $2/5$ & $1/3$ & $2/7$ & $1/4$ & $2/9$ & $1/5$ \\
\hline
$O_1(\alpha_n)$ & 1.31  & 1.22  & 1.16  & 1.13  & 1.11  & 1.09  & 1.08  & 1.07 \\
$O_2(\alpha_n)$ & 1.89  & 1.59  & 1.44  & 1.35  & 1.29  & 1.25  & 1.22  & 1.19 \\
$O_3(\alpha_n)$ & 2.31  & 1.85  & 1.62  & 1.49  & 1.40  & 1.34  & 1.30  & 1.26
\end{tabular}
\caption{
\label{overlap}
Overlap functions.}
\end{center}
\end{table}

Higher sensitivity to chameleons will be obtained with the next generation of experiments observing induced resonant transitions between two quantum states. 
In the first stage of GRANIT, the transition $3 \rightarrow 1$ of  neutrons in the terrestrial gravitational field  will be magnetically induced and the energy difference $E_3 - E_1$ will be measured with an estimated accuracy of $0.01$~peV.
In absence of exotic effects like the chameleon, the energy level expectation $E_3-E_1 = 1.91$~peV  is known very precisely from (\ref{solutionSchrodinger}).
Now the chameleon potential induces a shift $\delta E_i$ of the energy levels $E_i$ 
which can be calculated at  first order in perturbation theory:
\begin{equation}
\delta E_k = \left< \psi_k | \delta \Phi(z) | \psi_k \right>
\end{equation}
where $\delta \Phi(z) = \beta V_n (z/\lambda)^{\alpha_n}$ is a good approximation to the chameleon contribution to the potential for large $\beta$,  and $|\psi_k \rangle$ are the unperturbed wavefunctions.
We define the overlap functions:
\begin{equation}
O_k(\alpha) = \left< \psi_k \left| \left( \frac{z}{z_0} \right)^\alpha \right| \psi_k \right>.
\end{equation}
The overlap functions are calculated numerically for the first three quantum states, the result is shown in Table \ref{overlap}.
Then, the shift in the $3 \rightarrow 1$ resonance energy becomes
\begin{equation}
\label{GRANITsensitivity}
\delta E_{3-1} = \beta V_n \left( \frac{z_0}{\lambda} \right)^{\alpha_n} \left( O_3(\alpha_n) - O_1(\alpha_n) \right).
\end{equation}
In Fig. \ref{exclusion}, we  show the sensitivity of the first stage of GRANIT using eq. (\ref{GRANITsensitivity}), which is roughly independent of $n$: $\beta > 10^8$. 
This will greatly improve on the sensitivity of atomic experiments. 
We also show the ultimate sensitivity of the gravity-resonance-spectroscopy technique covering most of the chameleon parameter space. 

It is interesting to compare our results  to the limits set by macroscopic 5$^{th}$ force searches and equivalence principle (EP) tests
analysed in \cite{KhouryWeltman} and reproduced in fig. \ref{exclusion}.
Although these experiments are intrinsically sensitive to forces much weaker than gravity (especially for E.P. tests),
 chameleons can hide thanks to  the thin shell effect for large enough couplings $\beta$.
Especially, in all fifth force experiments measuring a force between two macroscopic bodies, a  shielding  layer
(to shield  magnetic, electric or thermal fluctuations) of thickness at least $10 \, \mu$m is inserted  between the two bodies.
This layer suppresses the chameleon-induced interaction for large enough couplings $\beta$.
The exclusion zone in \cite{KhouryWeltman} neglected this effect  which is  accounted for  in our fig. \ref{exclusion}.
Contrary to experiments using macroscopic bodies,  bouncing ultracold neutrons are free of any neutronic thin shell effect and allow one to
probe strongly coupled chameleons.
Strongly coupled chameleons could also be probed with Axion-like particle searches via the induced chameleon-photon coupling \cite{BraxBurrage}.
These experiments are sensitive to the coupling range $10^{11} < \beta < 10^{18}$ \cite{chase}.
For GRANIT, it turns out that the contribution from the magnetic energy to the density $\rho_\infty$  can be safely neglected.

During the preparation of this letter, the discovery of resonant transitions of the neutron quantum bouncer has been reported by the qBounce collaboration \cite{Jenke}. 
Although a detailed analysis of the systematic effects is not yet available, 
the $1 \rightarrow 3$ transition energy agrees with the Newtonian prediction at the level of $0.1$~peV. 
Thus the chameleon coupling to matter is constrained at the level of $\beta < 10^{9}$ already. 

We are grateful to A. Barrau and to members of the GRANIT collaboration (in particular D. Rebreyend, V.~Nesvizhevsky and K.~Protassov) for valuable discussions.

\bibliographystyle{unsrt}

\end{document}